\documentclass[12pt]{article}

\usepackage[yyyymmdd,hhmmss]{datetime}
\usepackage{amsmath,amssymb,amsfonts,amsthm}
\usepackage{authblk}
\usepackage{stackengine}
\usepackage{verbatim}
\usepackage{enumerate}
\usepackage{tikz}
\usepackage{algpseudocode}
\usepackage{algorithm}

\usepackage{url}
\PassOptionsToPackage{hyphens}{url}
\PassOptionsToPackage{hyphens}{hyperref}
\makeatletter
\g@addto@macro{\UrlBreaks}{\UrlOrds}
\makeatother

\usepackage{graphicx}

\allowdisplaybreaks

\usepackage[toc]{glossaries}

\usepackage{tocbibind}

\usepackage{lineno}

\usepackage{algorithm,algorithmicx,algpseudocode}



\floatname{algorithm}{Procedure}

\renewcommand{\algorithmicreturn}[1]{\bgroup\\  ~#1\egroup}
\renewcommand{\algorithmiccomment}[1]{\bgroup\hfill//~#1\egroup}

\theoremstyle{plain} \newtheorem{lemma}{\textbf{Lemma}}
\theoremstyle{plain} \newtheorem{proposition}{\textbf{Proposition}}
\theoremstyle{plain} \newtheorem{remark}{\textbf{Remark}}
\theoremstyle{plain} \newtheorem{theorem}{\textbf{Theorem}}
\theoremstyle{plain} 
\theoremstyle{plain} \newtheorem{example}{\textbf{Example}}
\theoremstyle{plain} 
\theoremstyle{plain} \newtheorem{property}{\textbf{Property}}
\theoremstyle{plain} \newtheorem{definition}{\textbf{Definition}}
\theoremstyle{definition}

\makeatletter
\newcommand{\pushright}[1]{\ifmeasuring@#1\else\omit\hfill$\displaystyle#1$\fi\ignorespaces}
\newcommand{\pushleft}[1]{\ifmeasuring@#1\else\omit$\displaystyle#1$\hfill\fi\ignorespaces}
\makeatother

\makeatletter
\let\@@pmod\pmod
\DeclareRobustCommand{\pmod}{\@ifstar\@pmods\@@pmod}
\def\@pmods#1{\mkern4mu({\operator@font mod}\mkern 6mu#1)}
\makeatother

\newcounter{NbCogito} 

\newcounter{NbFactum} 

\newcounter{NbTabulare} 

\newcounter{NbCoco} 

\newcommand{\rmv}[1]{}

\def\cA{{\mathcal A}}

\def\cK{{\mathcal K}}

\def\cB{{\mathcal B}}
\def\cC{{\mathcal C}}
\def\cE{{\mathcal E}}

\def\cL{{\mathcal L}}
\def\cN{{\mathcal N}}
\def\cR{{\mathcal R}}

\def\cW{{\mathcal W}}

\def\fb{{\mathfrak b}}
\def\fa{{\mathfrak a}}

\def\bF{{\mathbf F}}

\def\FF{{\mathbb F}}
\def\GG{{\mathbb G}}
\def\KK{{\mathbb K}}
\def\LL{{\mathbb L}}
\def\ZZ{{\mathbb Z}}

\def\TT{{\mathbb T}}

\def\rG{{\mathrm G}}
\def\rGF{{\mathrm {GF}}}
\def\rH{{\mathrm H}}

\def\rT{{\mathrm T}}

\providecommand{\myproofname}{Proof}

\title{On efficient normal bases over binary fields} 

\author{Mohamadou Sall}
\affil
{\stackunder{\small{Department of Electrical and Computer Engineering, University of Waterloo, Canada}}
{\mbox{\small{\texttt{msall@uwaterloo.ca}}}}
}

\author{M. Anwar Hasan}
\affil
{\stackunder{\small{Department of Electrical and Computer Engineering, University of Waterloo, Canada}}
{\mbox{\small{\texttt{ahasan@uwaterloo.ca}}}}
}

\date{}

\begin{document}

\maketitle
\thispagestyle{empty}

\vspace{-0.5cm}

\begin{abstract}
Binary field extensions are fundamental to many applications, such as multivariate public key cryptography, code-based cryptography, and error-correcting codes. Their implementation requires a foundation in number theory and algebraic geometry and necessitates the utilization of efficient bases.
The continuous increase in the power of computation, and the design of new (quantum) computers increase the threat to the security of systems and impose increasingly demanding encryption standards with huge polynomial or extension degrees. For cryptographic purposes or other common implementations of finite fields arithmetic, it is essential to explore a wide range of implementations with diverse bases. Unlike some bases, polynomial and Gaussian normal bases are well-documented and widely employed. In this paper, we explore other forms of bases of $\FF_{2^n}$ over $\FF_2$ to demonstrate efficient implementation of operations within different ranges. To achieve this, we leverage results on fast computations and elliptic periods introduced by Couveignes and Lercier, and subsequently expanded upon by Ezome and Sall. This leads to the establishment of new tables for efficient computation over binary fields.
\end{abstract}

\section{Introduction}

In numerous real-world applications a significant gap exists between fast computation and security level. Usually, for the sake of efficiency, designers of classical and post-quantum cryptosystems refrain from working in higher dimensions. Nonetheless,  there are situations where transitioning to those dimensions becomes necessary in order to improve the security of the designed schemes. For example, in the National Institute of Standards and Technology (NIST) call for post-quantum cryptography schemes there are three levels of security, and the complexity of computations increases from one level to the next. So, without specifying a particular cryptographic algorithm, it would be interesting to look for efficient implementation in large rings or finite fields. The goal of this paper is to delve into a deep analysis of efficient multiplications in large binary extensions for both common and extended implementations.

The approach is to exploit an explicit isomorphism between the binary field extension $\FF_{2^n}$ and the corresponding vector space $\FF_2^n$ through a basis that can be adapted to FFT (or NTT) like algorithms. Inherently the power basis $\{1, \ x, \ x^2, \ \cdots, \ x^{n-1}\}$ of $\FF_{2^n}$ over $\FF_2$ has such property and we can perform a multiplication using this basis in a quasi-linear time. Another type of basis that is usually encountered in the application of finite fields is the so-called normal basis \cite{Gao-Thesis}. A normal basis of $\FF_{2^n}$ over $\FF_2$ is a basis in the form $\{\alpha, \alpha^2, \cdots, \alpha^{2^{n-1}}\}$ where $\alpha \in \FF_{2^n}$ is a normal element. The simplest arithmetic operation in normal basis is squaring, carried out by a cyclic right shift, and hence almost free of cost in hardware. Such a cost advantage often makes normal basis a preferred
choice of representation, see work by Massey and Omura \cite{MO86} and its 
improvement by Reyhani and Hasan \cite{RH02, RH03}. Despite this main 
advantage, a normal basis multiplication is not so simple. So, we aim to
explore the ranges of normal bases that can be tied to NTT-like algorithms 
in order to perform fast multiplication in higher degrees $n$.

Optimal normal bases \cite{GSS07, MOVW89} and their generalization using Gauss periods \cite{ABV89} constitute an efficient way to implement finite fields arithmetic. Indeed, they usually have sparse {\it multiplication table} and as shown by Gao, von zur Gathen, Panario, and Shoup \cite{GGPS00} their construction can be tied to NTT-like algorithms. Nonetheless, for $q$ the cardinality of the finite field $\FF_q$, these bases use the multiplicative group of $\FF_{q^{r-1}}$ for some prime $r$ and do not always exist \cite{W90}. Later on,
taking advantage of the properties given by elliptic curves and other $1-$dimensional algebraic groups, Couveignes and Lercier \cite{CL09} and subsequently Ezome and Sall \cite{ES19} constructed {\it normal elements} that enlarge the number of cases where efficient normal bases exist. Indeed, these bases can also be tied to NTT-like algorithms and in some cases, we can get low complexity in term of number of non-zero elements of their multiplication tables. In a recent preprint Panario, Sall, and Wang \cite{PSW24} show that even if we may have the property of sparsity, it is very hard to predict the number of non-zero elements of the multiplication table of the bases designed by Couveignes and Lercier. Because of this fact, if multiplication tables are required for efficient implementation, we will be mainly focused on other types of bases.

Gaussian normal bases (GNBs) have been studied for decades; they are well understood and there is a lot of computational material available online \cite{HFFWeb, PackGit} featuring their properties. This is not the case for
normal bases from elliptic periods or extension of normal elements. Indeed most of the data available in the literature are mainly based on results from GNBs and trace computation (see \cite[Table 2.2.10]{MP13}). 

From the resolution of Fermat's last theorem by Andrew Wiles to their recent utilization in post-quantum cryptography (through isogenies), elliptic curves, and by extension algebraic groups, have always given an interesting turn in the research area. In this paper, we use their properties to establish new
ranges for fast computation. We make extensive use of the results described in 
\cite{CL09, ES19} and the extended normal elements developed by Thomson and Weir 
\cite{TW18} and generalized to the Artin-Schreier-Witt and Kummer theories by Ezome and Sall \cite{ES20}. We should notice that the elliptic normal bases
of $\FF_{q^n}$ over $\FF_q$ designed by Couveignes and Lercier require an appropriate elliptic curve $E(\FF_q)$ whose existence is related to an integer $n_q$. This integer is implicitly tied to the number of $\FF_q-$rational  points of $E$ and should satisfy the condition $n_q\leq \sqrt{q}$. This is just too large and can be very tight for a binary extension. Indeed it is quasi-impossible to find an
appropriate curve over $\FF_2$. Even if workarounds are presented in \cite{CL09} and \cite{CE23}, the developed strategies involve either many ad hoc redundant representations or lack of a convenient model for a universal curve of genus $g$ over $\FF_q$, which make it difficult to handle in practice. In this paper, we show how to use practical elliptic normal bases over binary fields by exploiting the characteristics of central processing unit (CPU) registers and the design of computational software.
The data, we presented in various tables show how the use of elliptic curves 
can enlarge the number of cases where efficient normal bases exist.

The paper is organized as follows. In Section \ref{sec:BNTTA} we recall the results of fast computation using NTT-like algorithms. In Section \ref{sec:NBGP} we give the necessary backgrounds on Gaussian normal bases. Our efforts culminate in Section \ref{sec:NBAG} and Section \ref{sec:CUSNE} where we give new tables for fast computations over binary fields.
In each section, we highlight how the concerned basis is tied to NTT-like algorithms. Finally, in Section \ref{sec:CR}, we give concluding remarks.
Our studies are supported by Magma codes that we have developed and made available 
online.


\section{Background on NTT algorithms}\label{sec:BNTTA}

The search for algorithms with fast computation in a ring or finite field extension is an important research area. This problem can be approached according to different models of construction and representation. For each computational problem, one should examine whether the type of representation for the objects (e.g., polynomial, integer) at hand is of use. For the sake of efficiency, one may ask: In which representation is the problem easiest to solve? How do we convert to and from that representation? The Discrete Fourier Transform and its implementation, the Fast Fourier Transform (FFT), are the backbone of the
general idea of change of representation.\\
\indent Let $\cR$ be a ring, $n\geq 1$ an integer, and $\zeta\in \cR$ be a 
primitive $n$-th root of unity. Let $f\in \cR[X]$ be a polynomial of degree less than $n$ with coefficient vector $(f_0, \cdots, f_{n-1})\in \cR^n$.
\begin{definition}
The $\cR-$linear map $$f \longmapsto \left(f(1), \cdots, f(\zeta^{n-1})\right)$$ which evaluates a polynomial at the powers of $\zeta$ is called the Discrete Fourier Transform. The inverse operation is the interpolation at the powers 
of $\zeta$.
\end{definition}

An important result in computational algebra is that both the DFT and its inverse can be computed with $O(n\log n)$ operations in $\cR$.
However, if the underlying coefficient ring $\cR$ is arbitrary, an FFT-based approach will not work directly. Indeed, this approach works over rings containing certain primitive roots of unity. As a counter-example, a finite field $\FF_q$ contains a primitive $n$-th root of unity
if and only if $n$ divides $q-1$. In 1971, Schonhage and Strassen \cite{SS71} showed how to create ``virtual" roots that leads to a multiplication cost of only $O(n\log n\log\log n)$.
This idea gives rise to many studies and improvements, and recently Harvey and Hoeven \cite{HH19} proved that multiplication in $\FF_{q^n}$ can be done with $O(n\log n)$ operations in $\FF_q$.\\
\indent In order to achieve quasi-linear speed in $O(n\log n)$ when performing polynomial multiplication or any related operation (e.g., convolution product of vectors) we use the FFT to compute the Discrete Fourier Transform or more specifically the Number Theoretic Transform (NTT) \cite{P71}.  DFT and NTT are
just different names for the same thing depending on the context.
Indeed, NTTs are DFTs defined in finite fields and rings of polynomials or integers. Their main advantage over DFTs is that systematic advantage is taken from the operation in finite fields or rings to define simple roots of unity which allows to replacement of complex arithmetic by real arithmetic. The general principle of NTT can be found in many algorithmic books because it is widely used in many engineering, signal processing, mathematical and physical domains.

\begin{definition}
In this paper, an NTT-like algorithm is an algorithm that involves the computation of polynomials of degree less than $n$ or convolution products of vectors of length $n$.
\end{definition}

Even though NTT-like methods are the backbone of the fastest algorithms,
general-purpose computer algebra systems also implement the classical and sometimes the Karatsuba method and its variants, which are sub-quadratic. So, to perform efficient polynomial or vector arithmetic a large variety of algorithms have to be coded and tested to determine the crossover point. That is the point where one algorithm beats another in terms of performance (running time, number of operations, memory size).
For small degree or length $n$, the school book method is sufficient, and the Karatsuba algorithm is used when we have intermediate $n$. If the integer $n$ is very large, it becomes much more interesting to work with NTT methods.

The choice of degree for using the Number Theoretic Transform (NTT) in the context of binary extension fields typically involves several considerations.
Generally, the Karatsuba algorithm and NTT excel under different operand size 
ranges. Finding precise crossover points can be empirical and is often dependent on hardware architecture, implementation details, and the specific characteristics of the multiplication tasks. For sizes less than a few hundred bits, the Karatsuba algorithm might perform better and for larger operand sizes, e.g., several hundred bits or more, the NTT might outperform. In binary fields, the bit sizes are directly related to the degrees $n$ of the extensions. This motivates us to work with different ranges for $n$ starting from 250. It is noteworthy that in some contexts the users can limit themselves to other algorithms like Karatsuba's one and can utilize our codes in \cite{PackGit} to customize the values according to their needs.


\section{Normal bases from Gauss periods}\label{sec:NBGP}

An early interest in normal bases may stem from the fact that Gauss 
\cite{GaussDA} used normal elements to show how to draw a regular polygon 
with ruler and compass alone. This problem remained unsolved for more than 2000 years and he gave its solution using elements that he called periods. 
Gauss' insights laid the foundation for many subsequent developments in
theory as well as in practice. Indeed, after approximately 200 years, in 
1989 Ash, Blake, and Vanstone \cite{ABV89} used these same periods to 
investigate low complexity normal bases over finite fields.

\begin{definition}
Let $q$ be a power of a prime number $p$. Let $r=nk+1$ be a prime number not 
dividing $q$ and $\gamma$ a primitive $r$-th root of unity in $\FF_{q^{nk}}$. 
Let $K$ be the unique subgroup of order $k$ of $\ZZ_{r}^*$ and $\cK_i\subseteq \ZZ_{r}^* $ be a coset of $K$ for $0\leq i \leq n-1$. The elements
$$\alpha_i= \sum_{a\in \cK_i}\gamma^a \in \FF_{q^n}, \ 0\leq i \leq n-1$$
are called Gauss period of type $(n,k)$ over $\FF_q$.
\end{definition}

After 10 years of Ash et al. \cite{ABV89} findings, Gao, Gathen, Panario, and Shoup \cite{GGPS00} showed how  Gaussian normal bases of $\FF_{q^n}$ over $\FF_q$ can be adapted to NTT-like algorithms. In the following, we highlight their strategy.
Let $\cN_{gp} = (\alpha_0, \alpha_1, \cdots, \alpha_{n-1})$ be a normal basis
of $\FF_{q^n}$ over $\FF_q$ generated by a Gauss period $\alpha_0$ of type $(n, k)$. Let  $$A=\sum_{0\leq i\leq n-1}a_i\alpha_i, \ B=\sum_{0\leq i\leq n-1}b_i\alpha_i, \text{ and } C = \sum_{0\leq i\leq n-1}c_i\alpha_i$$
the product of $A$ and $B$ computed in $\cN_{gp}$ representation. The equation
$$A=\sum_{0\leq i\leq n-1}a_i\sum_{j\in \cK_i}\gamma^j = \sum_{1\leq j\leq nk}a_j'$$
leads us to computations using polynomials of degree $\leq nk$. Indeed, 
let $$\Phi_r = x^r-1$$ be the $r$ cyclotomic polynomial, $\cL= \FF_q[X]/(\Phi_r)$ 
the quotient ring and $x= X \mod \Phi_r$. Then $\cL$ has two power bases
$$\{1, \ x, \ \cdots, \ x^{nk-1}\} \text{ and } \{x, \ x^2, \ \cdots, \ x^{nk}\}$$
over $\FF_q$. The advantage of this change of representation is that it is
easy to go from one basis to another and the elements of $\cL$ can be viewed as
polynomials of degree less than $nk-1$ or $nk$. The map 
$$\varphi: \FF_{q^n} \longrightarrow \cL, \ A \longmapsto \sum_{1\leq j\leq nk}a_j'x^j $$ is a ring homomorphism with image $\varphi(\FF_{q^n})$ given by the
subring $$\cL_1 = \{ c_1x+c_2x+\cdots+c_{nk}x^{nk} \in \cL \ : \ c_1, \cdots, 
c_{nk} \in \FF_q \text{ and } c_i=c_j \text{ for } i, \ j \in \cK_i\}$$
For every nonzero element $A$ of $\FF_{q^n}$, $\varphi(A)$ can be easily
inverted. So, to compute $C$, we multiply $\varphi(A)$ and $\varphi(B)$,
perform the reduction in $\cL$ and then invert the result back to $\FF_{q^n}$. 
{\it This is how Gaussian normal bases are tied to NTT-like algorithms}. The basic 
idea is to convert the normal basis representation to a polynomial basis in the 
ring $\cL$, do fast multiplication of polynomials in the ring, and then convert 
the result back to the normal basis. We give a summary of the result in the
following theorem, for more details see \cite[Section 4]{GGPS00}.

\bigskip

\begin{theorem}[\cite{GGPS00}]\label{theo:GNBmult}
Let $\cN$ be a normal basis generated by a Gauss period of type $(n, k)$ over $\FF_q$. Then  multiplication in $\cN$ can be computed with $O(nk \log(nk)\log\log(nk))$ operations in $\FF_q$.
\end{theorem}

A normal basis generated by a Gauss period of type $(n, k)$ is often simply called Gaussian normal basis (GNB) of type $k$.
Theorem \ref{theo:GNBmult} shows that it is very important to work with GNBs of small
type. The parameter $k$ should be kept as small as possible not only 
because of the cost of multiplication but also for the structure of the multiplication
table $\rT_{\cN}$ of $\cN$. Indeed, as stated in Proposition \ref{Prop:GNBcomplexity}, in binary extensions the sparsity of $\rT_{\cN}$ depends on $k$.

\begin{proposition}[\cite{ABV89, BGM91, CGPT12}]\label{Prop:GNBcomplexity} 
    Let $n>3$ be an integer and $\cN$ be a normal arising from a Gauss period of type $(n, k) $ for $\FF_{2^n}$ over $\FF_2$. Let $\cC_{\cN}$ be the complexity of $\cN$ in term of number of nonzero elements of its multiplication table $\rT_{\cN}$.
    
    \begin{enumerate}
        \item We have the following bounds on $\cC_{\cN}$.
        $$
           \begin{cases}
               kn - (k^2 - 3k + 3) \leq \cC_{\cN} \leq (n - 1)k + 1 \ \ \ \ \ \ \ \ \ \ \ \ \ \text{ if } k  \text{ is even,}\\
               (k + 1)n - (k^2 - k + 1) \leq \cC_{\cN} \leq (n-2)k + n + 1 \ \text{ if } k \ \text{ is odd.}
          \end{cases}
         $$
         \item If $1\leq k\leq 6$ then we have the following exact complexities:
         \begin{itemize}
             \item If $k=1$ or $2$ then $\cN$ is optimal and $\cC_{\cN}=2n-1$.
             \item If $k=3$ or $4$ then $\cC_{\cN}=4n-7$.
             \item If $k=5$ or $6$ then $\cC_{\cN}=6n-21$.
         \end{itemize}
    \end{enumerate}
\end{proposition}

In practice, it is recommended to choose the type $k$ with the lowest achievable complexity. So, in this paper, we work with Gauss periods of type no more than 10. However, the reader should note that the algorithms (e.g., the Magma functions {\it MultNBrange} and {\it ExtNBrange}) we developed for the data in Sections \ref{sec:NBAG} and \ref{sec:CUSNE} can be extended to higher types $k$. We conclude this section by giving a table of GNBs whose lowest types are less or equal to $10$ for $\FF_{2^n}$ over $\FF_2$, where $250\leq n \leq 600$. The range $$250\leq n \leq 600$$ is of particular interest for common implementations and we need the properties (cost of multiplication, complexity) of GNBs in this range to work with sub-normal elements in Section \ref{sec:CUSNE}. Some more data related to Table \ref{tab:GNB250_600} are available in \cite{HFFWeb, PackGit}.

\begin{table}
    \centering
    \begin{tabular}{| c | c | c | c | c | c | c | c | c | c | c |}
    \hline
         $(n, k)$ & $(n, k)$  & $(n, k)$ & $(n, k)$ & $(n, k)$ & $(n, k)$ & $(n, k)$ & $(n, k)$ & $(n, k)$ \\
    \hline   
    250, 9 & 284, 3 & 326, 2 & 364, 3  & 407, 8  & 446, 6  & 491, 2 & 532, 3 & 572, 5 \\
    251, 2 &  285, 10 &  327, 8 & 367, 6 & 409, 4  & 447, 6  & 493, 4  & 534, 7 & 573, 4 \\
    252, 3 &  286, 3 &  329, 2 & 369, 10 & 410, 2  & 449, 8  & 494, 3  & 535, 4 & 574, 3  \\
    253, 10 &  287, 6 & 330, 2 & 370, 6  & 411, 2  & 451, 6  & 495, 2 & 537, 8  & 575, 2 \\ 
    254, 2 & 290, 5  & 331, 6 & 371, 2  & 412, 3  & 453, 2  & 498, 9 & 538, 6  & 577, 4 \\
    255, 6 &  291, 6 & 332, 3 & 372, 1  & 413, 2  & 458, 6  & 499, 4 & 540, 1  & 578, 6  \\ 
    257, 6 &  292, 1 & 334, 7 & 373, 4  & 414, 2  & 459, 8  & 501, 10 & 542, 3  & 579, 10  \\  
    258, 5 &  293, 2 & 337, 10 & 374, 3 & 417, 4  & 460, 1 & 502, 10 & 543, 2  & 580, 3  \\ 
    259, 10 &  294, 3 & 338, 2 & 375, 2 & 418, 1  & 461, 6 & 503, 6 & 545, 2  & 581, 8 \\
    260, 5 & 297, 6  & 339, 8 & 378, 1 & 419, 2  & 462, 10  & 505, 10 & 546, 1  & 582, 3 \\ 
    261, 2 &  298, 6 & 340, 3 & 380, 5 & 420, 1  & 465, 4  & 506, 1 & 547, 10  & 583, 4 \\
    262, 3 &  299, 2 & 341, 8 & 381, 8  & 421, 10  & 466, 1  & 507, 4 & 548, 5  & 585, 2 \\
    263, 6 &  301, 10 & 342, 6 & 382, 6 & 423, 4  & 467, 6  & 508, 1 & 550, 7  & 586, 1  \\ 
    265, 4 &  302, 3 & 343, 4 & 385, 6 & 425, 6  & 469, 4  & 509, 2 & 551, 6  & 589, 4 \\ 
    266, 6 &  303, 2 & 345, 4 & 386, 2  & 426, 2  & 470, 2  & 510, 3 & 553, 4  & 591, 6 \\ 
    267, 8 &  305, 6 & 346, 1 & 387, 4  & 428, 5  & 471, 8  & 511, 6 & 554, 2 & 593, 2 \\ 
    268, 1 & 306, 2 & 347, 6 & 388, 1  & 429, 2  & 473, 2  & 513, 4 & 555, 4  & 595, 6 \\ 
    269, 8 & 307, 4 & 348, 1 & 390, 3  & 430, 3  & 474, 5  & 515, 2 & 556, 1  & 596, 3 \\     
    270, 2 &  310, 6 & 349, 10 & 391, 6  & 431, 2  & 475, 4  & 516, 3 & 557, 6  & 597, 4 \\ 
    271, 6 &  311, 6 & 350, 2 & 393, 2  & 433, 4  & 476, 5  & 517, 4 & 558, 2  & 599, 8 \\ 
    273, 2 &  313, 6 & 351, 10 & 394, 9  & 434, 9  & 478, 7 & 519, 2 & 559, 4  &    \\ 
    274, 9 &  314, 5 & 354, 2 & 395, 6  & 435, 4  & 479, 8  & 522, 1 & 561, 2  &  \\
    276, 3 &  315, 8 & 355, 6 & 397, 6  & 438, 2  & 481, 6  & 523, 10 & 562, 1  &  \\
    277, 4 &  316, 1 & 356, 3 & 398, 2  & 439, 10  & 482, 5  & 524, 5 & 564, 3  &  \\
    278, 2 &  319, 4 & 357, 10 & 401, 8  & 441, 2  & 483, 2  & 525, 8 & 565, 10  &   \\ 
    279, 4 & 322, 6 &  358, 10 & 402, 5  & 442, 1  & 484, 3  & 526, 3 & 566, 3  &  \\ 
    281, 2 & 323, 2 & 359, 2 & 404, 3  & 443, 2  & 486, 10  & 527, 6 & 567, 4  &  \\ 
    282, 6 & 324, 5  & 362, 5 & 405, 4  & 444, 5  & 487, 4  & 530, 2 & 570, 5  &   \\ 
    283, 6 & 325, 4  & 363, 4 &406, 6  & 445, 6  & 490, 1  & 531, 2 & 571, 10  &  \\
    \hline         
    \end{tabular}
    \caption{GNBs of type $\leq10$ for $\FF_{2^n}$ over $\FF_2$, where $250\leq n \leq 600$.}
    \label{tab:GNB250_600}
\end{table}


\section{Normal bases from algebraic groups}\label{sec:NBAG}

After 10 years of Gao et al.'s work on fast computation using Gauss periods \cite{GGPS00}, Couveignes and Lercier \cite{CL09} presented an alternative approach to the construction of efficient normal bases by introducing elliptic periods. These bases leverage the properties of elliptic curves, making it easier to expand the range of cases where they exist. Another 10 years later, Ezome and Sall \cite{ES19} applied a similar strategy to all remaining $1-$dimensional algebraic groups over a finite field $\FF_q$.

\begin{proposition}[\cite{M06}, Chapter II, Section 3]\label{prop:4AGs}
    Any $1-$dimensional connected algebraic group over a perfect field $\KK$ is isomorphic to one of the following:
    \begin{enumerate}
        \item an elliptic curve group $E$,
        \item an additive group $\GG_a$,
        \item multiplicative groupe $\GG_m$, or
        \item Lucas torus $\TT_{\alpha}:x^2-\alpha y^2=1$
        
    \end{enumerate}
\end{proposition}

We recall that any finite field is a perfect field. The following theorem summarizes all the results on normal bases and fast multiplication derived from $1-$dimensional algebraic groups.

\begin{theorem}[\cite{CL09, ES19}]\label{theo:CL_ES}
    Let $\rG$ be a $1-$dimensional algebraic group over $\FF_q$ and $t\in\rG$ a point
    of order $n$. Let $\rH=\rG/(t)$ and $I:\rG \longrightarrow \rH$ the isogegy quotient.
    Let $a$ be a point of $\rH(\FF_q)$ such that $I^{-1}(a)$ is irreducible over $\FF_q$. Then $\FF_q(I^{-1}(a))$ is a cyclic extension of $\FF_q$ of degree $n$. Let $\cL$ be a linear space given by functions over $\Bar{\FF}_q(\rG)$. Assume there is a point $R\in \rG(\FF_q)$ such that the application $$f \longmapsto (f(R+jt))_{0\leq j\leq n-1}$$ is a bijection from $\cL$ to $\FF_q^n$. Then there is a normal basis $\cN_{ag}$ of $\FF_{q^n}$ over $\FF_q$ with NTT-like multiplication algorithm. 
\end{theorem}

In the algebraic approach, first of all, we need to construct an extension $\LL$ of degree $n$ of a field $\KK$ using the properties of an appropriate algebraic 
group $\rG$. Note that in this section, $\LL$ is equal to the finite field $\FF_{q^n}$.
The geometric origin of this extension results in a nice description
of $\KK$-automorphisms of $\LL$. Indeed, let $b\in I^{-1}(a)$ and $O$ the identity 
element of $\rG$. Let $(*_G)$ be the group law of $(\rG, *_G)$ and by restriction $(*_H)$ be the law in the subgroup $\rH$. Since
$$I(t*_Gb) = I(t)*_HI(b) = O*_Ha = a$$ 
then $t*_Gb$ is Galois  conjugated to $b$ and all conjugates are obtained that way from all $n-$torsion points $(kt)_{0\leq k\leq n-1}$ in the subgroup $\rT$ of order $n$ generated by $t$. So we have an isomorphism    
$$\psi: t \longmapsto \varphi_t$$  between $\rT$ and the Galois group of $\LL$ over $\KK$, where
$\varphi_t$ is the residual automorphism $$b\in I^{-1}(a) \longmapsto b*_Gt.$$
In the context of the four algebraic groups enumerated in Proposition \ref{prop:4AGs}, 
the geometric formulas that describe the translation $x \mapsto x*_Gt$ in 
$\rG$ give a nice description of the Galois group of $\LL$ over $\KK$, which
leads to a succinct construction of $\cN_{ag}$. For more details see \cite{CL09, ES19}.\\
\indent The linear space $\cL$ of dimension $n$ plays a central role in the computation
over $\cN_{ag}$. It results from the theory of divisors over algebraic groups.
For example, we can define $\cL$ as follows:
$$\cL = \{ f\in \Bar{\KK}(\rG) \ : \ div(f) \geq -D \}\cup\{0\}$$
where $div(f)$ represents the divisor of any function $f$ over $\rG$ and 
$$D = \sum_{0\leq k\leq n-1} k(t) $$
is the divisor given by the formal sum of the points in $\rT$. For more
details on the theory of divisors we refer to \cite[Chapter 2, Section 3]{Silv09}. The main advantage of $\cL$ is that it allows a bijective correspondence 
$$\cL \longleftrightarrow \FF_q^n\longleftrightarrow \FF_{q^n}$$

\noindent between functions on $\rG$ and scalars in $\FF_q$ given by a representation
in the normal basis $\cN_{ag}$ of $\FF_{q^n}$ over $\FF_q$. 
This correspondence is possible because $\cN_{ag}$ is obtained by evaluating
the elements of a basis $\cB$ of $\cL$ at the $\FF_{q^n}-$rational point 
$b\in I^{-1}(a)$. Even if we are in different algebraic settings, the reader can note the similarity between the importance of the linear space $\cL$ and the ring $\cL$ used in the Gaussian normal basis fast multiplication algorithm. This is why we gave the same notation.

The element $R$ is an $\FF_q-$rational point of $\rG$ and as outlined in Theorem \ref{theo:CL_ES} it is core to fast multiplication in a normal basis from algebraic group. The multiplication tensor of $\cN_{ag}$ is based on evaluation and interpolation, and involves additions, subtractions, component-wise products, and convolution products
between vectors of length $n$. The first three operations are very easy to perform, and the latter one can be seen as a polynomial
multiplication. {\it This is how normal bases from algebraic groups (NBAGs) are tied to NTT-like algorithms}. For more details in the study of fast algorithms and convolution products, we refer to \cite{GG03} and to see how it is applied to normal basis construction, we refer to \cite{CL09, ES19}.

Each algebraic group mentioned in Proposition \ref{prop:4AGs} possesses unique strengths and limitations. The Lucas Torus, specifically, operates exclusively within finite fields where the characteristic is $p\neq2$. As such, since we are in the context of binary field extensions it falls outside the scope of our study.
Normal bases derived from both additive and multiplicative groups offer the 
additional benefit of featuring a sparse multiplication table. Nonetheless, 
they are particularly well-suited for Artin-Schreier and Kummer extensions.

The main advantage of elliptic normal bases (ENBs) over GNBs and other normal
bases from algebraic groups is that for a given finite field $\FF_q$ we can
define many elliptic curves. This characteristic significantly expands the range of cases where a normal basis, employing an NTT-like algorithm, can exist. However, it's crucial to note that implementing an elliptic normal basis is a complex task. Furthermore, as demonstrated in the following lemma (derived from Theorem \ref{theo:CL_ES}), the degree $n$ should not be excessively large.

\begin{lemma}[\cite{CL09}]\label{the:main:theo:couv-lerc}
Let $p$ be a prime number, $q$ a power of $p$, and $n\geq2$ an integer.
Let $\ell$ be any prime number and $v_{\ell}$ the valuation associated
to $\ell$. We denote by $n_q$ the unique positive integer satisfying
\begin{itemize}
    \item $v_{\ell}(n_q)=v_{\ell}(n)$ if  $\ell$ is coprime with  $q-1$,
	\item $v_{\ell}(n_q)=0$ if $v_{\ell}(n)=0$, 
	\item $v_{\ell}(n_q)=\mathrm{max}\{2v_{\ell}(q-1)+1, 2v_{\ell}(n)\}$ if $\ell | q-1$ and $\ell | n$.
\end{itemize}
If $n_q\leq \sqrt{q}$, then there exists an elliptic curve $E$ over $\FF_q$ with
a point $t$ of order $n$, a point $b$ in $E(\FF_{q^n})$, and a $\FF_q-$rational
point $R$ outside of $E[n]$. These points allow the construction of an elliptic normal basis with NTT-like multiplication algorithms.
\end{lemma}

The assumption $n_q\leq \sqrt{q}$ poses a significant constraint in applications involving binary field extensions $\FF_{2^n}/\FF_2$, primarily due to the inherent difficulty of finding a suitable elliptic curve over $\mathbb{F}_2$. Although a workaround is proposed in \cite{CL09}, the devised strategy introduces numerous ad hoc redundant representations, making practical implementation challenging. In the subsequent sections, we leverage the architecture and software design of computers to demonstrate an efficient implementation of elliptic normal bases on binary field extensions.

\subsection{Choice of good embedding degrees}

In the realm of finite field arithmetic, the efficiency of various operations, such as addition and multiplication relies on the chosen representation for the elements. While Zech logarithms could be applicable for small-sized fields, these methods become impractical for larger finite fields commonly employed in modern public key cryptography. Consequently, it is customary to leverage an explicit isomorphism between the field $\FF_{q^n}$ and the corresponding vector space $\FF_{q}^n$, achieved through a carefully selected basis.\\
\indent As obtaining a straightforward isomorphism proves challenging within the framework of elliptic normal bases of $\FF_{2^n}$ over $\FF_2$, this section capitalizes on the implementation of Zech logarithms, exploits the characteristics of computer registers, and utilizes the embedding of finite fields. These measures are employed to enable better functionality of elliptic normal bases over binary extensions. The basic idea is based on the following scheme
\begin{center}

\begin{tikzpicture}[node distance=2cm, auto]
  \node (F_2) {$\FF_2$};
  \node (F_2e) [right of=F_2] {$\FF_{2^e}$}; 
  \node (F_2n) [right of=F_2e] {$\FF_{2^n}$}; 
  \draw[-latex] (F_2) -- (F_2e) node[midway] {$e$};
  \draw[-latex] (F_2e) -- (F_2n) node[midway] {$d$};
\end{tikzpicture}
\end{center}

\noindent where $\FF_{2^e}$ is a subfield of $\FF_{2^n}$, and the degrees of the extensions
$\FF_{2^n}/\FF_{2^e}$ and $\FF_{2^e}/\FF_{2}$ are respectively equal to $d$ and $e$.
To perform operations (e.g., multiplication) in $\FF_{2^n}$ we split the computations
in two stages. Initially, we focus on identifying a suitable normal basis for 
$\FF_{2^n}/\FF_{2^e}$ by leveraging algebraic group properties. We keep the degree $e$ as small as possible so that the computations in $\FF_{2^e}/\FF_{2}$ will be very easy to perform. For example, we can choose $e$ such that multiplication and addition can be done through a Zech logarithms table. The goal is to have the smallest embedding degree $e$ while ensuring $n_q\leq \sqrt{q}$, where $q=2^e$. So, a natural question is how to upper bound $e$? We will do that based on computer architecture and some software design.

In CPUs and digital systems, the execution of operations within binary finite fields is commonly achieved through the utilization of registers and combinational logic circuits. Registers, serving as high speed data storage elements in processors, store information on which combinational logic circuits perform operations. For instance, in the case of addition or XOR operations, the contents of two registers are XORed bit by bit to generate the sum.

Multiplication, on the other hand, entails a sequence of XOR and bit-shift operations. The multiplicand and multiplier are loaded into registers, and the product is accumulated through a series of XOR and shift operations. The final result undergoes reduction modulo an irreducible polynomial. The implementation of these operations relies on fundamental components of digital circuits, including XOR gates, AND gates, OR gates, and shift registers.

This underscores the importance of choosing embedding degrees of $2^k$, typically with $k$ values such as 4, 5, or 6, corresponding to register sizes of 16, 32, and 64. Opting for smaller data sizes, like 16 bits, reduces the memory footprint of data structures and variables compared to larger sizes such as 32 or 64 bits. Moreover, in real-time systems where timely and predictable responses are crucial, using smaller data sizes can enhance execution speed by processing smaller data chunks in each operation. To meet the requirement of facilitating easy computations in $\FF_{2^e}$, it is advisable to select embedding degrees no greater than 16.

Beyond considerations of CPU micro-architecture, the design of certain software can further support the rationale behind opting for a modest embedding degree. For instance, within the Magma system \cite{MAS}, arithmetic operations in small non-prime finite fields are facilitated through Zech logarithm tables. When the field $\FF_{2^e}$ is sufficiently compact to be represented using Zech logarithms, the output of elements is formatted as powers of a primitive element that generates $\FF_{2^e}$. 
The following Magma code 
\begin{verbatim}
    for e in [2..20] do
         F := GF(2^e); AssertAttribute(F, "PowerPrinting", true); 
         print Random(F);
    end for;
\end{verbatim}
prints the elements of $\FF_{2^e}$, for $2\leq e\leq20$ in power mode. However, if we go beyond $20$ even if the function {\it AssertAttribute} is set to `true', the current version of the Magma system returns the following statement
\begin{verbatim}
     Runtime error in `AssertAttribute': Power printing is not 
     available for this kind of field
\end{verbatim}
and starts representing elements using polynomials. This means at this point we can't use Zech logarithms for field representation. So taking this fact into account we can also choose a small embedding degree $e\leq20$.

By synthesizing the two justifications outlined in the preceding paragraphs, this paper focuses on small embedding degrees $e\leq20$. It's worth noting that, upon reviewing Table \ref{tab:ENBtab}, most extensions $\FF_{2^n}/\FF_2$ exhibit embedding degrees of 16 or less. As mentioned earlier, the value 16 holds particular significance due to its alignment with CPU micro-architecture. To illustrate the devised strategy clearly, we provide the following example.

\begin{example}
Let $n=507$, using the divisors of $n$, we get the following diagram.
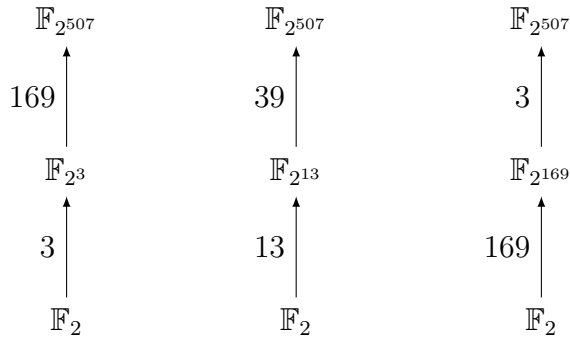
\begin{center}
\begin{tikzpicture}[node distance=2cm, auto]
  \node (F_2) {$\FF_2$};
  \node (F_2e) [above of=F_2] {$\FF_{2^{3}}$}; 
  \node (F_2n) [above of=F_2e] {$\FF_{2^{507}}$}; 
  \draw[-latex] (F_2) -- (F_2e) node[midway] {$3$};
  \draw[-latex] (F_2e) -- (F_2n) node[midway] {$169$};
\end{tikzpicture} \hspace{1.5cm}
\begin{tikzpicture}[node distance=2cm, auto]
  \node (F_2) {$\FF_2$};
  \node (F_2e) [above of=F_2] {$\FF_{2^{13}}$}; 
  \node (F_2n) [above of=F_2e] {$\FF_{2^{507}}$}; 
  \draw[-latex] (F_2) -- (F_2e) node[midway] {$13$};
  \draw[-latex] (F_2e) -- (F_2n) node[midway] {$39$};
\end{tikzpicture} \hspace{1.5cm}
\begin{tikzpicture}[node distance=2cm, auto]
  \node (F_2) {$\FF_2$};
  \node (F_2e) [above of=F_2] {$\FF_{2^{169}}$}; 
  \node (F_2n) [above of=F_2e] {$\FF_{2^{507}}$}; 
  \draw[-latex] (F_2) -- (F_2e) node[midway] {$169$};
  \draw[-latex] (F_2e) -- (F_2n) node[midway] {$3$};
\end{tikzpicture}
\end{center}

The smallest embedding degree $e$ for which $n_q\leq\sqrt{q}$, where $q=2^e$, is $13$.
For a counterexample, there is no embedding degree $e\leq20$ satisfying the condition $n_q\leq\sqrt{q}$ if $n=500$. That is why this value is missing in Table \ref{tab:ENBtab}.
\end{example}

\subsection{Table for efficient computations}

The data table presented in this section primarily relies on normal bases derived from elliptic curves. Given the close correlation between the existence of a suitable curve and the integer $n_q$ as defined in Lemma \ref{the:main:theo:couv-lerc}, it becomes imperative to comprehend its inherent properties.

\begin{property}\label{property:n_q}
Let $p$ be a prime, $q$ a power of $p$. Let $n\geq2$ and $e\geq1$ be two integers.
We denote by $\varphi$ the Euler function.
\begin{enumerate}
    \item If $n$ is prime to $q-1$ then $n_q=n$.
    \item If $e$ is prime to $n\varphi(n)$ then $n_q=n_{q^e}$.
    \item  If $q-1$ is squarefree then $n_q\leq n^3$.
    \item  In any case $n_q\leq n^2(q-1)^2$.
\end{enumerate}
\end{property}

The algorithms, as described in \cite{PackGit}, that lead to the creation of Table \ref{tab:ENBtab} are notably influenced by the parameter $n_q$. To provide a tangible perspective on this value, we offer the following example featuring two instances of the pair $(n, q)$. 

\begin{example}
The initial instance involves a finite field with a characteristic distinct from $2$, while the second instance pertains to binary extensions.
\begin{enumerate}
    \item We consider the finite field extension $\FF_{q^n}$ where $n=10$ and $q =42989$. Since $n=2\times 5$ then $2$ and $5$ are the only primes $\ell$ that appear in the decomposition of $n_q$. So computing $n_q$ amounts to finding
    $v_2(n_q)$ and $v_5(n_q)$. We have the following factorization of 
     $$q-1 = 2^2\times11\times977.$$ Since $2$ divides both $n$ and $q-1$, by the third point of Lemma \ref{the:main:theo:couv-lerc} $v_2(n_q) = 5$. Since $5$ is prime to $q-1$, by the first point of this lemma $v_5(n_q) = v_5(n)=1$. So finally, we have $$n_q = 2^5\times 5 = 160.$$
     \item We consider the binary field extension $\FF_{2^{976}}$ over $\FF_{2^{16}}$. So, for this instance $q=2^{16}$ and the degree of the extension $n=61$. Since 
     $$gcd(61, \ 2^{16}-1)=1,$$ by Point (1) of Property \ref{property:n_q},
     $n_q=n=61$. In this case, we can construct an efficient elliptic normal basis because we have the sufficient condition $n_q\leq \sqrt{q}$.
\end{enumerate}
\end{example}

We can now compile Table \ref{tab:ENBtab}, presenting a comprehensive list of elliptic normal bases with embedding degrees $embed\leq20$ within the range $500\leq n\leq 1000$. This particular range has been selected to accommodate highly demanding applications. However, utilizing our algorithms detailed in \cite{PackGit}, readers can easily customize the values and generate a table tailored to their specific requirements.

\begin{center}
\begin{table}
    \centering
    \begin{tabular}{| c | c ||| c | c ||| c | c ||| c | c ||| c | c |}
    \hline
         $n$ & $embed$  & $n$ & $embed$ & $n$ & $embed$ & $n$ & $embed$ & $n$ & $embed$\\
    \hline         
         507 & 13 & 611 & 13 & 714 & 17 & 817 & 19 & 931 & 19 \\ 
         510 & 15 & 612 & 17 & 715 & 13 & 819 & 13 & 935 & 17 \\
         512 & 16 & 615 & 15 & 720 & 15 & 825 & 15 & 936 & 13 \\
         513 & 19 & 616 & 14 & 722 & 19 & 826 & 14 & 938 & 14\\
         516 & 12 & 624 & 13 & 728 & 13 & 828 & 18 & 940 & 20 \\        
         518 & 14 & 627 & 19 & 731 & 17 & 832 & 13 & 944 & 16\\       
         520 & 13 & 629 & 17 & 732 & 12 & 833 & 17 & 949 & 13 \\
         522 & 18 & 636 & 12 & 736 & 16 & 836 & 19 & 950 & 19\\ 
         527 & 17 & 637 & 13 & 738 & 18 & 840 & 20 & 952 & 14\\
         528 & 12 & 640 & 20 & 740 & 20 & 845 & 13 & 954 & 18\\ 
         532 & 14 & 644 & 14 & 741 & 13 & 846 & 18 & 960 & 15\\
         533 & 13 & 645 & 15 & 742 & 14 & 848 & 16 & 962 & 13 \\
         540 & 15 & 646 & 17 & 744 & 12 & 850 & 17 & 969 & 17 \\
         544 & 17 & 650 & 13 & 748 & 17 & 854 & 14 & 975 & 13\\
         546 & 13 & 656 & 16 & 750 & 15 & 855 & 15 & 976 & 16 \\
         551 & 19 & 658 & 14 & 752 & 16 & 858 & 13 & 980 & 14\\
         552 & 12 & 660 & 15 & 754 & 13 & 860 & 20 & 986 & 17 \\
         555 & 15 & 663 & 13 & 760 & 19 & 867 & 17 & 988 & 13\\
         558 & 18 & 665 & 19 & 765 & 15 & 868 & 14 & 990 & 15 \\
         559 & 13 & 666 & 18 & 767 & 13 & 870 & 15 & 992 & 16\\         
         560 & 14 & 675 & 15 & 768 & 12 & 871 & 13 & 994 & 14\\
         561 & 17 & 676 & 13 & 770 & 14 & 874 & 19  &  &  \\
         564 & 12 & 680 & 17 & 774 & 18 & 884 & 13  &  & \\
         570 & 15 & 684 & 19 & 779 & 19 & 885 & 15  &  & \\
         572 & 13 & 686 & 14 & 780 & 13 & 893 & 19  &  &  \\
         574 & 14 & 688 & 16 & 782 & 17 & 896 & 14  &  &  \\         
         576 & 18 & 689 & 13 & 784 & 14 & 897 & 13  &  &  \\
         578 & 17 & 690 & 15 & 792 & 18 & 900 & 15  &  & \\
         580 & 20 & 696 & 12 & 793 & 13 & 901 & 17  &  &  \\
         585 & 13 & 697 & 17 & 795 & 15 & 910 & 13  &  &  \\
         589 & 19 & 700 & 14 & 798 & 19 & 912 & 19  &  & \\
         592 & 16 & 702 & 13 & 799 & 17 & 915 & 15  &  &  \\
         595 & 17 & 703 & 19 & 806 & 13 & 918 & 17  &  & \\
         598 & 13 & 704 & 16 & 810 & 15 & 920 & 20  &  & \\
         600 & 15 & 705 & 15 & 812 & 14 & 923 & 13  &  & \\
         608 & 16 & 708 & 12 & 816 & 17 & 928 & 16  &  &  \\
         %

    \hline         
    \end{tabular}
    \caption{$\FF_{2^n}$ with ENB of embedding degree $\leq 20$ for $500\leq n\leq 1000$.}
    \label{tab:ENBtab}
\end{table}
\end{center}

\begin{remark}
    In Table \ref{tab:ENBtab}, our primary emphasis lies on elliptic normal bases due to their expansive range of possibilities compared to other bases. However, a normal basis derived from the multiplicative group, aside from being easily implementable, can exhibit superior characteristics when it is available. For instance, our Magma function, named {\it MultNBrange} \cite{PackGit}, executed within the range $500\leq n \leq 1000$, yields embedding degrees of $9$ and $11$ for respective values of $657$ and $979$. Although such cases are relatively scarce, they seem to offer improved computational efficiency when encountered. Note that
    there is no ENB with embedding degree $\leq20$ when $n$ is equal to $657$ or $979$.
\end{remark}


\section{Computation using sub-normal elements}\label{sec:CUSNE}

Polynomial bases and normal bases from Gauss periods and $1-$dimensional algebraic groups are not the only bases of $\FF_{q^n}$ over $\FF_q$ that allow fast multiplication using NTT-like algorithms. Indeed, as shown by Thomson and Weir 
\cite{TW18} and then Ezome and Sall \cite{ES20}, we can take advantage of Artin-Schreier-Witt and Kummer theory to construct bases that are not normal basis but can exploit the advantageous properties of normal elements. For binary extensions, 
these bases are obtained by extending (if possible) an appropriate normal basis 
of $\FF_{2^{n/2}}, \ \FF_{2^{n/3}},$ or $\FF_{2^{n/4}}$ over $\FF_2$.\\
\indent The goal of this section is to compile a table that reports the integers 
$n$ for which a GNB of type less than 10 does not exist. As a workaround, we 
precise the feasibility of working with normal element extensions or elliptic normal bases with small embedding degrees. Before delving into this analysis, it is crucial to comprehend the conditions under which a Gauss period of type $(n, k)$ over $\FF_q$ generates a normal basis. This inquiry is generally addressed by the Wasserman condition \cite{W90}. We present the refined version for binary extensions.

\begin{theorem}\label{the:Wass_cond}
    There exists a Gaussian normal basis of $\FF_{2^n}$ over $\FF_2$ if and only if $8$ does not divide $n$.
\end{theorem}
 
Utilizing Theorem \ref{the:Wass_cond} and algorithms developed in \cite{ES20, TW18}, we can broaden the scope of cases where multiplication in $\FF_{2^n}/\FF_2$ can be efficiently performed through NTT-like algorithms and normal elements.
The basic idea is as follows: if there are no efficient normal bases of $\FF_{2^n}/\FF_2$, one may hope that $n$ has a proper divisor $d$ such that $\FF_{2^d}/\FF_2$  admits an efficient normal basis $\cN=\{\alpha, \alpha^2, \cdots, \alpha^{2^{d-1}}\}$. 
 We consider the cases where $d$ is chosen to be either $n/2$, $n/4$, or $n/3$, aligning with values that are respectively effective in Artin-Schreir theory and its generalization to Witt vectors and Kummer theory. 
The following theorem gives a summary of the result derived from the first two theories. For background on Artin-Schreir theory and Witt vectors, we refer the reader to \cite{Wit36} and \cite[Section 2.2]{ES20}.

\begin{theorem}[\cite{ES20, TW18}]\label{the:AWalg}
Let $\cN=\{\alpha, \alpha^2, \cdots, \alpha^{2^{d-1}}\}$ be a normal basis of $\FF_{2^d}/\FF_2$. Then there exists $\fa\in \FF_{2^{2d}}$ and $\fb_0$ and $\fb_1$ in $\FF_{2^{4d}}$ such that $$\cA=\cN \cup \fa\cN$$ is a basis of $\FF_{2^{2d}}$ over $\FF_2$ define as a degree 2 Artin-Schreier extension of $\cN$
and $$\mathcal{W}=(\mathcal{N}\cup \fb_0 \mathcal{N})\cup \fb_1(\mathcal{N}\cup \fb_0\mathcal{N})$$ is a basis of $\FF_{2^{4d}}$ over $\FF_2$ define as a degree 4 Artin-Schreier-Witt extension of $\cN$. If $d$ is odd we can directly construct
$\mathcal{W}$ as a degree 2 extension of $\cA$.
\begin{enumerate}
\item The cost of multiplication in $\cA$ consists in at most 3 multiplications and 4 additions between elements lying in $\FF_{2^d}$ and 1 vector-matrix multiplication between a vector of $\FF_{2^d}/\FF_2$ and the multiplication table of $\cN$.
\item The cost of multiplication in $\cW$ consists in at most 9 multiplications, 33 additions between elements lying in $\FF_{2^d},$ and 9 vector-matrix multiplications between a vector of $\FF_2^d$ and the multiplication table of $\cN$.
\item  If $\cN$ has subquadratic cost of multiplication and subquadratic complexity, then $\cA$ and $\cW$ have also sub-quadratic cost of multiplication.
\end{enumerate}
\end{theorem}

Let $A\in\FF_{2^{2d}}$, then $A$ can be represented in the extended basis $\cA$ as
$$A=A_0+\fa A_1, \text{ where } A_0 = \sum_{i=0}^{d-1}a_{0i}\alpha^{2^i} \text{ and }
A_1 = \sum_{j=0}^{d-1}a_{1j}\alpha^{2^j}.$$
We can represent an element $B$ of $\FF_{2^{4d}}$ in $\cW$ in a similar way. {\it So, 
it is clear that if $\cN$ supports an NNT-like algorithm, then so does $\cA$ and 
$\cW$, and a natural question is how to choose $\cN$?}

\indent Theorem \ref{the:AWalg} shows that it is important to find a normal basis $\cN$ with NTT-like algorithm and such that its multiplication table $\rT_{\cN}$ is as sparse as possible. This property is featured by normal bases from additive and multiplicative groups and according to Theorem \ref{theo:GNBmult} and Proposition 
\ref{Prop:GNBcomplexity} Gauss periods of small type $(d, k)$ are also good 
candidates. Unfortunately, in terms of multiplication with an NTT-like method, the 
first two bases are not adapted to extensions of the form $\FF_{2^d}/\FF_2$.
For example, as shown in \cite[Theorem 1]{ES19}, in order to tie a normal basis
from the additive group to NTT-like algorithms the base field $\FF_p$ should be a proper subfield of $\FF_q$, the underlying field of $\FF_{q^n}$. This
requirement is impossible in the situation of this section. On the other hand, elliptic normal bases have fast multiplication algorithms but it is hard to expect a low complexity from them (see the recent preprint by Panario, Sall, and Wang \cite{PSW24}). Consequently, to the best of our knowledge, Gaussian normal bases of
low types are the best candidates to work with Artin-Schreier-Witt extensions of 
$\FF_{2^n}$ over $\FF_2$.

To establish Table \ref{tab:1T_2T} we work with the range $1000\leq n\leq 1200$
in order to go beyond the possibilities given in Table \ref{tab:ENBtab}. However, the interested readers can customize the values and generate a table tailored to their specific requirements using the algorithms available in \cite{PackGit}. In this section, for the given
range, we report the integers $n$ for which there is no GNB of type $\leq 10$. Next,  for $d\in \{n/2, \ n/3, \ n/4\}$ we check if $\FF_{2^d}/\FF_2$ has a GNB of type $k\leq 10$. In the table, if possible, $(d, k)$ is defined for each of the three types of extended bases: $\cA, \ \cW,$ and $\cK$. The Kummer basis extensions, defined by the column $\cK$, are obtained from Theorem \ref{theo:KumExt}. Note that these extensions work for $d=n/3$ and contrary to Artin-Schreir-Witt extensions we need the additional condition $3 \ | \ 2^n-1$.  Finally as an additional study, in the column $\cE$, we check if an elliptic normal basis with embedding degree $\leq20$ exists for the extension $\FF_{2^n}/\FF_2$. If such an $\cE$ exists we note the corresponding embedding degree, 
if not we leave the field blank.

\begin{theorem}[\cite{ES20}]\label{theo:KumExt}
Let $d$ be a positive integer such that $3$ divides $2^d-1$. Assume that $\cN = \{\alpha, \alpha^2, \cdots, \alpha^{2^{d-1}}\}$ is a primitive normal basis $\FF_{2^d}/\FF_2$. Then there exists $\beta$ in $\bF_{2^{3d}}$ such that $\Omega=\cN\cup\beta\cN\cup\beta^2\cN$ is a degree $3$ Kummer extension 
of $\cN$.
\begin{enumerate}
\item The complexity of $\Omega$ consists of at most 6 multiplications and 15 additions between elements of $\FF_{2^d}$, 2 vector-matrix multiplications between vectors in $\FF_2^d$  and the multiplication table of $\cN$.
\item  If $\cN$ has subquadratic complexity and subquadratic weight in $d$, then $\Omega$ has also subquadratic complexity in $d$. 
\end{enumerate}
\end{theorem}

\noindent The following Magma code 
\begin{verbatim}
    KL := [339, 349, 351, 358, 359, 364, 381, 387, 397];
    for d in KL do
        if (2^d-1) mod 3 eq 0 then print d; end if;
    end for;
\end{verbatim}
returns only 358 and 364, which means the integers $d$ satisfying the additional condition $3$ divides $2^d-1$ correspond respectively to the values $n=1074$ and $n=1092$ in Table \ref{tab:1T_2T}. Furthermore, in the case of Kummer extensions  we may need a primitive normal element and it is shown in \cite{LS87} such element exists for any finite field extension $\FF_{q^n}/\FF_q$. Note that looking for a concrete normal element and assessing its properties (e.g., being primitive or efficient or having low complexity) can be resource-demanding \cite{MPT18} and is out of the scope of our study. However, we can read the following conjecture in \cite{GGP98}. {\it For any positive integer $d$ not divisible by 8, there exists an integer $k > 1$ such that the Gauss period of type $(d, k)$ is primitive normal in $\FF_{2^n}$ over $\FF_2$}. An experimental study indicates that this is true for all $d<569$. Except the five cases where $d\in\{ 571, \ 572, \ 575, \ 580, \ 596 \}$ all the underlying Gaussian normal bases in Table \ref{tab:1T_2T} satisfy this range. Since the excepted values do not concern Kummer extensions, there is a high probability of getting an appropriate primitive Gaussian normal basis that works well for our study.

\begin{center}
\begin{table}
    \centering
    \begin{tabular}{|c|c|c|c|c|||c|c|c|c|c|}
    \hline
         $n$ & $\cA$  & $\cW$  & $\cK$ & $\cE$ & $n$ & $\cA$ & $\cW$ & $\cK$ & $\cE$  \\
    \hline       
   
         1000 &  & (250, 9) &  &  & 1109 &   &   &  &   \\ 
         \hline
         1007 &  &  &  & 19 & 1111 &   &   &   &  \\ 
         \hline
         1008 &  & (252, 3) &  &  & 1112 & (556, 1) & (278, 2) &   &  \\
         \hline
         1016 & (508, 1) & (254, 2) &  &  & 1114 & (557, 6) &   &   &  \\
         \hline
         1017 &  &  & (339, 8)  &  & 1120 &   &  &  & 14  \\ 
         \hline
         1024 &  &  &  & 16  & 1128 & (564, 3) & (282, 6) &   &  \\
         \hline
         1028 &  & (257, 6) &  &  & 1132 & (566, 3) & (283, 6) &   &  \\ 
         \hline
         1032 & (516, 3) & (258, 5) &  &  & 1136 &   & (284, 3) &   & 16  \\
         \hline
         1040 &  &  (260, 5) &  & 13  & 1139 &   &  &  & 17  \\ 
         \hline
         1042 &  &  &  &  & 1141 &   &   &  &   \\ 
         \hline
         1047 &  &  & (349, 10)  &  & 1142 & (571, 10) &   &   &  \\          
         \hline
         1048 & (524, 5) & (262, 3) &  &  & 1143 &  &   & (381, 8) &   \\ 
         \hline
         1051 &  &  &  &  & 1144 & (572, 5) & (286, 3) &   & 13   \\ 
         \hline
         1053 &  &  &  (351, 10) & 13  & 1149 &  &   &   & \\
         \hline
         1056 &  &  &  &  & 1150 & (575, 2) &   &   &  \\          
         \hline
         1058 &  &  &  &  & 1152 &   &   &   & 18   \\ 
         \hline
         1059 &  &  &  &  & 1153 &   &  &   &    \\
         \hline
         1064 & (532, 3) & (266, 6) &   & 14  & 1160 & (580, 3) & (290, 5) &   & 20 \\        
         \hline
         1072 &  &  (268, 1) &  & 16  & 1161 &   &  & (387, 4) &  \\ 
         \hline
         1073 &  &  &  &  & 1163 &   &  &   &  \\
         \hline
         1074 & (537, 8) &  &  (358, 10)  &  & 1168 &   & (292, 1) &   & 16 \\
         \hline
         1077 &  &  & (359, 2)  &  & 1175 &   &  &   &  \\
         \hline
         1079 &  &  &  & 13  & 1176 &   & (294, 3) &   & \\
         \hline
         1080 & (540, 1) & (270, 2) &  &  15  & 1177 &  &  &   &  \\
         \hline
         1081 &  &  &  &  & 1180 &   &  &   & 20 \\
         \hline
         1085 &  &  &  &  & 1181 &   &  &   &  \\
         \hline
         1088 &  &  &  & 17  & 1184 &   &  &   & 16\\
         \hline
         1092 & (546, 1) & (273, 2) &  (364, 3)  & 13  & 1188 &  & (297, 6) &  & \\
         \hline
         1094 & (547, 10) &  &  &  & 1189 &   &  &   & \\
         \hline
         1095 &  &  &  &  15  & 1191 &   &  & (397, 6) & \\
         \hline
         1096 & (548, 5) & (274, 9) &  &  & 1192 & (596, 3) & (298, 6) &  & \\
         \hline
         1097 &   &  &  &  & 1195 &   &  &   &  \\ 
         \hline
         1104 &   & (276, 3) &  &  & 1196 &   & (299, 2) &   & \\
         \hline
         1105 &   &  &  & 13  & 1200 &   &  &   & 15 \\         
    \hline         
    \end{tabular}
    \caption{Extensions $\FF_{2^n}/\FF_2$ without GNB of type $\leq10$ for $1000\leq n\leq 1200$.}
    \label{tab:1T_2T}
\end{table}
\end{center}

\section{Concluding remarks}\label{sec:CR}

It is clear that the strategies described in Section \ref{sec:NBAG} and \ref{sec:CUSNE} do not work if $n$ is a prime number $p$. Fortunately, we still have the possibility of using efficient normal bases. Indeed, as shown by Theorem \ref{the:Wass_cond}, a Gaussian normal basis that can be adapted to NTT-like algorithms exists for the binary extension $\FF_{2^p}/\FF_2$. The only
remaining question is how to choose the type $k$ of this basis, which depends on the user's needs. If any GNB of type $k$ over $\FF_{2^p}$ does not satisfy the user's expectations, another possibilty is to embed $\FF_{2^p}$ itself into a field $\FF_{2^{ep}}$ using the following scheme:

\begin{center}
\begin{tikzpicture}[node distance=2cm, auto]
  \node (F_2) {$\FF_2$};
  \node (F_2e) [right of=F_2] {$\FF_{2^p}$}; 
  \node (F_2n) [right of=F_2e] {$\FF_{2^{ep}}$.}; 
  \draw[-latex] (F_2) -- (F_2e) node[midway] {$p$};
  \draw[-latex] (F_2e) -- (F_2n) node[midway] {$e$};
\end{tikzpicture}
\end{center}

\noindent Next, we check if $\FF_{2^{ep}}$ over $\FF_2$ admits a GNB of type $k$
that satifies our requirements. A list of GNBs of lowest type $(n, k)$ for $n\leq2000$ is available in \cite{HFFWeb, PackGit}. In this case, since the elements of $\FF_{2^p}$ are interpreted as elements of $\FF_{2^{2p}}$ we will lose (slightly) computational efficiency, and it is better to choose $e=2$ or as small as possible.


\end{document}